\documentclass[useAMS,usenatbib,usegraphicx]{mn2e}

\title[Dynamically-Driven Star Formation In Models Of NGC~7252]{Dynamically-Driven Star Formation In Models Of NGC~7252}
\author[L.-H. Chien and J. E. Barnes]{L.-H. Chien$^{1}$\thanks{E-mail:
chien@stsci.edu (LHC); barnes@ifa.hawaii.edu (JEB)} and 
J. E. Barnes$^{2}$\\
$^{1}$Space Telescope Science Institute,\\
3700 San Martin Drive, Baltimore, MD 21218, USA\\
$^{2}$Institute for Astronomy, University of Hawaii\\
2680 Woodlawn Drive, Honolulu, HI 96822, USA
}

\newcommand{\dd}{{\it density-dependent} }
\newcommand{\si}{{\it shock-induced} }
\newcommand{\idkit}{{\it Identikit} }

\begin{document}
\maketitle

\begin{abstract}
We present new dynamical models of the merger remnant NGC~7252 which include star formation simulated according to various phenomenological rules.  By using interactive software to match our model with the observed morphology and gas velocity field, we obtain a consistent dynamical model for NGC~7252.  In our models, this proto-elliptical galaxy formed by the merger of two similar gas-rich disk galaxies which fell together with an initial pericentric separation of $\sim2$ disk scale lengths approximately $620$~Myr ago.  Results from two different star formation rules--- density-dependent and shock-induced--- show significant differences in star formation during and after the first passage.  Shock-induced star formation yields a prompt and wide-spread starburst at the time of first passage, while density-dependent star formation predicts a more slowly rising and centrally concentrated starburst.  A comparison of the distributions and ages of observed clusters with results of our simulations favors shock-induced mechanism of star formation in NGC~7252.  We also present simulated color images of our model of NGC~7252, constructed by incorporating population synthesis with radiative transfer and dust attenuation.  Overall the predicted magnitudes and colors of the models are consistent with observations, although the simulated tails are fainter and redder than observed.  We suggest that a lack of star formation in the tails, reflected by the redder colors, is due to an incomplete description of star formation in our models rather than insufficient gas in the tails.
\end{abstract}


\section{Introduction}
NGC~7252 is a well-studied merger remnant \citep{too77,s83}, which provides strong observational evidence supporting the idea that at least some disk galaxy mergers produce elliptical galaxies.  Its luminosity profile follows a $r^{1/4}$ law \citep{s82}, as do the profiles of elliptical galaxies \citep{deV, kor}.  The core contains a central counter-rotating disk of ionized gas within $8\arcsec$ ($\simeq2.5$~kpc\footnote{In order to compare with previous observational results, throughout this paper we adopt $H_{0}=100h=75$~km~s$^{-1}$~Mpc$^{-1}$.  This places NGC~7252 at a distance of 64.4~Mpc and a projected scale of $1\arcsec=$~312~pc.}) of the center, providing a kinematic signature of a major merger event in the recent past \citep{s82}.  Molecular gas is also closely associated with the central disk \citep{dup,wang}.  The two tidal tails are rich in H\,{\small I}, suggesting that the progenitors were late-type spirals \citep[][hereafter HGvGS94]{hb94}.  About $500$ candidate young clusters are found in the main body in the {\em Hubble Space Telescope} images \citep{whit,m97}, indicating that clusters can form in large numbers in the violent star formation that accompanies a galaxy merger, and that cluster formation can increase the specific globular cluster frequency of the merger remnant.  Moreover, since the clusters are single burst stellar populations, the gravitationally bound clusters surviving several $10^{8}$ years provide a record of the merger's star formation history.  Ages and metallicities of these clusters can be used to reconstruct that history; \citet[hereafter SS98]{ss} found five young globular clusters with ages of $400$--$600$~Myr, which presumably formed shortly after the first close encounter of the two gas-rich disk galaxies that later merged to produce the present-day remnant.  Metallicities of these clusters support the idea that elliptical galaxies with bimodal globular cluster systems can form through major mergers.

In addition to the extensive observations of NGC~7252, many numerical models have been made to investigate the history of the system.  \citet[hereafter HM95]{hm} revised encounter parameters from previous studies \citep{br, m93} with detailed H\,{\small I} observations from the Very Large Array (VLA, HGvGS94).  Their dynamical simulation determined the age of NGC~7252 to be about $773$ ($\sim580h^{-1}$)~Myr after periapse of the initial orbit and predicted the future infall rate of the tidal tail materials.  \citet{mdh} further constrained the dark halos in the NGC~7252 progenitor galaxies.  These models matched the morphology and kinematics of the system very well, and have provided robust results for the dynamical history.   However, gas dynamics and star formation have not yet been included in any dynamical modeling of NGC~7252.  \citet{ag} attempted to reconstruct NGC~7252's star formation history and spectro-photometric properties using an evolutionary spectral synthesis model, but were forced to postulate that the interaction-induced star formation began well before the first passage.

In this paper we use simulations including gas dynamics and star formation to construct possible star formation histories for NGC~7252.  We follow \citet{b04} in examining how two different star formation rules--- \dd and $shock$-$induced$--- affect the star formation history.  In addition to matching the observed morphology and H\,{\small I} kinematics, we compare our star formation models with the ages of young globular clusters observed by SS98.  We examine the different star formation models and try to constrain the triggers of star formation in merging galaxies.  Moreover, we also compare the modeled colors of NGC~7252 with the observed values.  Our goal is to match the overall morphology, kinematics and photometric properties while insuring that the initial disks have reasonable parameters for late-type spirals.  This represents a new and significant step beyond pure dynamical modeling.  This paper is organized as follows: in Section~2 we describe the methodology and the simulation technique, while in Section~3 we describe the matching process to the H\,{\small I} observation as well as our best-fit model.  In Section~4 we present the predicted star formation history and a comparison with the ages of SS98 clusters.   We discuss the simulated images of NGC~7252 and a comparison of the photometric properties in Section 5, and finally give conclusions in Section 6.  For the purpose of a consistent discussion throughout the paper, we introduce physical scales in Section 2 using the scaling factors of the best-fit model, which will be described in Section~3.

\section{Numerical Techniques}
\subsection{Galaxy Models}
\label{galmodel}
We use bulge/disk/halo galaxy models similar to those developed by \citet[][Appendix B]{bh}.  Each galaxy has a bulge with a shallow cusp \citep{hq}, an exponential disk with constant scale height, and a truncated dark matter halo with a Navarro-Frenk-White profile \citep{n97}.  The mass and scale lengths of each component is listed in Table~\ref{comparam}.  The stellar disk's $Q$ parameter \citep{too64} is initially greater than 1.45 at all radii, insuring local stability and helping to suppress spontaneous bar formation.  The total mass of the disk is $M_{disk}\simeq1.87\times10^{10}$~M$_{\sun}$, slightly smaller than the Milky Way but reasonable for typical spiral galaxies.  The disk has a circular velocity $v_{c}(R=3R_{disk})\simeq176$~km~s$^{-1}$.  We also estimate its maximum rotation velocity, $v_{max}$, from the Tully-Fisher relation \citep{tf}, assuming that the total mass, $M$, is proportional to the fourth power of the circular velocity, or $M\propto v^{4}_{c}$.  Using the Milky Way as a reference (i.e.~a disk mass of $M_{MW}=6.0\times10^{10}$~M$_{\sun}$\ and circular velocity of 220~km~s$^{-1}$), we predict a circular velocity for our galaxy model $v_{max}=(M_{disk}/M_{MW})^{1/4} \times 220 =164$~km~s$^{-1}$.  This value is slightly smaller than the circular velocity of our model galaxy, but well within the observed scatter of the Tully-Fisher relation.  

\begin{table}
   \centering
   \caption{Parameters of the components of each progenitor galaxy.}
   \label{comparam}
   \begin{tabular}{@{}cc}
   \hline
   Component & Value\\
   \hline
   M$_{bulge}$ & 0.62$\times10^{10}$~M$_{\sun}$\\
   M$_{disk,\,stellar}$ & 1.40$\times10^{10}$~M$_{\sun}$\\
   M$_{disk,\,gas}$ & 0.47$\times10^{10}$~M$_{\sun}$\\
   M$_{halo}$ & 9.96$\times10^{10}$~M$_{\sun}$\\
   a$_{bulge}$$^{a}$ & 0.44~kpc\\
   z$_{disk,\,stellar}$$^{b}$ & 0.28~kpc\\
   z$_{disk,\,gas}$$^{b}$ & 0.11~kpc\\
   R$_{disk}$$^{c}$ & 1.84~kpc\\
   a$_{halo}$$^{d}$ & 5.53~kpc\\
   \hline
   \end{tabular}\\
  \medskip
  $^{a}$Bulge scale length.\\
   $^{b}$Disk scale height.\\
   $^{c}$Disk radial scale length.\\
   $^{d}$Halo scale radius.\\
\end{table}

The total halo to disk-plus-bulge mass ratio of our galaxy is 4:1.  Thus our galaxy models have relatively low-mass halos compared to those expected in some scenarios of galaxy formation in CDM \citep[e.g.][]{mmw98}, although they are not very different from 
the models used in previous simulations of NGC~7252.  We conjecture that the orbital decay of a pair of galaxies following a relatively close passage is not very sensitive to the outer halo structure; this could be explored in experiments embedding galaxy encounters in a self-consistent cosmological context, but it is beyond the scope of the present study.  

Gas is initially distributed like the stellar disk and follows an isothermal equation of state with temperature $T\sim10^{4}$~K.  The star formation history depends on the amount of gas that is available in both galaxies before the encounter.  SS98 suggested that NGC~7252 is formed by the merger of two gas-rich disk galaxies, possibly of Hubble type Sc.  Accordingly, we made the progenitor galaxies gas rich, with gas comprising $25\%$ of the disk mass.  The total number of particles in each galaxy is $108544$, broken down into $8192$ bulge particles, $18432$ disk particles, $32768$ halo particles, and $49152$ gas particles, which are evolved using a smoothed particle hydrodynamics (SPH) code.  Gravitational forces calculated by a modified N-body tree code with a Plummer smoothing of $\epsilon\simeq0.17$~kpc.  

\subsection{Star Formation Prescriptions}
The star formation rules incorporated in our simulation follow \citet{b04}.  We adopt the star formation prescription:
\begin{equation}
\label{sfp}
\dot{\rho}_{\ast} \propto C_{\ast}~\rho_{g}^{n}~\dot{u}^{m},
\end{equation}
\noindent where the star formation rate, $\dot{\rho}_{\ast}$, depends on both gas density, $\rho_{g}$, and local rate of mechanical heating due to shocks, $\dot{u}$.  This prescription is implemented in a probabilistic fashion \citep{b04}.  At each time step, the probability of a gas particle undergoing star formation is calculated according to the assumed star formation rule.  If a gas particle satisfies the probabilistic criteria for star formation, it is instantly and completely transformed into a stellar particle.  This scheme provides a Monte-Carlo description of star formation-- consistent with the overall Monte-Carlo nature of N-body simulation-- while keeping the total number of particles fixed and avoiding complications associated with composite star/gas particles.  For later analysis, each stellar particle is tagged with the value of time at its birth.  

We considered two limiting regimes for NGC~7252: \dd and \si star formation.  The former star formation model is solely depending on the accumulation of gas, with $n=1.5$ and $m=0$ mimicking a Schmidt law \citep{schmidt,kenni98}.  The latter model uses $n=1.0$ and $m=0.5$, favoring star formation in regions corresponding to shocks.  As described in \citet{b04}, the constant $C_{\ast}$ has to be adjusted to obtain equivalent results in simulations with different spatial resolutions, and to insure that different simulations consume about the same amount of gas. We adopt $C_{\ast}=0.018$ for the \dd model and $C_{\ast}=0.5$ for the \si model.

\subsection{Star Formation Prehistory\label{sfph}}
\begin{figure}
  \centering
   \includegraphics[height=3.3in,angle=-90]{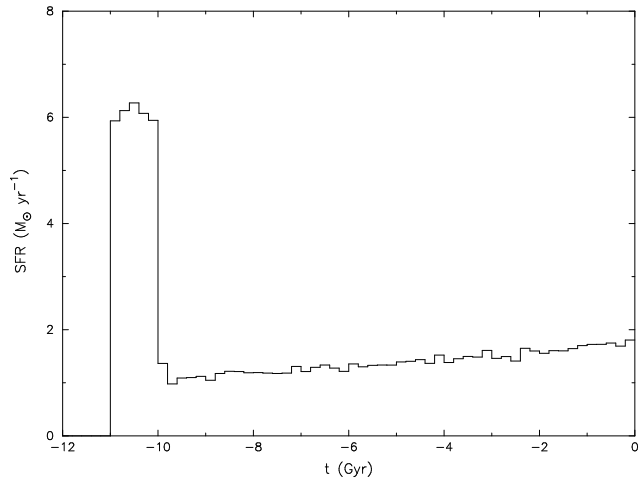}
   \caption{Star formation prehistory of the progenitor galaxies  (in M$_{\sun}$~yr$^{-1}$) as a function of time (in Gyr); $t=0$ is when the simulation starts.  Disk populations began forming 10 Gyr ago and bulge populations formed 1 Gyr before the disk, corresponding to the spike.  The small fluctuations in the star formation rate reflects our monte-carlo implementation of star formation.}
   \label{presf}
\end{figure}

In the previous section we describe how we model ongoing star formation, but in order to calculate the photometric properties, we also need to specify the pre-encounter star formation history of the initial galaxies.  We setup the pre-encounter star formation history of the initial galaxies as shown in Figure~\ref{presf}.  First, we assume the bulge formed between 11 and 10~Gyr before the start of the simulation ($t$ = 0); since the integrated photometric properties of such an old population are insensitive to its detailed star formation history, we adopt a constant star formation rate (SFR) of 6.2~M$_{\sun}$~yr$^{-1}$ during bulge formation.  Second, we assume the disk began forming 10~Gyr before the start of the simulation and adopt an exponential form for the disk's SFR \citep{searle73,s86},
\begin{equation}
\label{sfh}
\Psi(t)=\dot{M}_{0}\, e^{-\alpha\,t}, 
\end{equation}
\noindent where $\dot{M}_{0}$ ($\simeq1.8$~M$_{\sun}$~yr$^{-1}$) matches the dynamically-driven SFR at the start of the simulation.  To determine the $e$-folding factor, $\alpha$, we integrate equation (\ref{sfh}) from $t=-10$~Gyr to $t=0$ and require the result to be equal to the mass of the stellar disk.  In our simulation, each of the progenitor galaxies has a stellar disk mass of $1.40\times10^{10}$~M$_{\sun}$, so this factor is found to be $\alpha\simeq-0.054$~Gyr$^{-1}$.  The SFR thus rises slowly throughout the previous 10~Gyr, with an average of 1.4~M$_{\sun}$~yr$^{-1}$.  This gives the characterization parameter $b$ \citep{scalo86}, defined as the ratio of the current SFR to the past SFR averaged over the age of the disk, a value of about 1.3, consistent with typical late-type spirals.  We emphasize that this star formation history is schematic; it is broadly consistent with known constraints, but other scenarios are possible.  For example, instead of using a slightly unconventional e-folding factor $\alpha < 0$, we could have given the disk a constant SFR of 1.8~M$_{\sun}$~yr$^{-1}$ and a maximum age of 7.7~Gyr; this would make the the pre-encounter disks a bit more luminous but would have very little effect on the overall outcome of our modeling.  
  
Finally, the birth-dates of individual bulge and disk particles are assigned using a Monte-Carlo procedure.  Bulge particle birth-dates are drawn from a uniform distribution between $-11$ and $-10$~Gyr, reflecting the constant SFR of the bulge assumed above.  Disk particle birth-dates are assigned by sampling the disk SFR $\Psi(t)$ defined by equation (2).

\subsection{Modeling Photometric Properties\label{mphot}}
The photometric properties of the modeled galaxies are determined by convolving the simulation output with a simple stellar population model.  Using the conventional notation for SPH, the luminosity density, $j$, in waveband $X$, can be expressed as:
\begin{equation}
j_{X}({\bf r},t)=\sum_{i} W({\bf r}-{\bf r}_{i}, h) \cdot \ell_{X}(t-t_{i}),
\end{equation}
\noindent where $W$ is the kernel function, $h$ is the smoothing length, and $\ell_{X}(\tau)$ is the luminosity of a stellar particle with age $\tau$.  The sum runs over all stellar particles; particle $i$ has position {\bf r$_{i}$} and birthdate $t_{i}$.

To evaluate particle luminosities, we use \citet[][hereafter BC03]{bc} models for a stellar population with solar metallicity and \citet{ch03}  initial mass function (IMF).  One somewhat subtle issue concerns the mass returned to the interstellar medium (ISM) via winds, planetary nebulae, and supernova ejecta.  Following the normal population synthesis methodology, BC03 tabulate luminosities as functions of time for a stellar population formed in a burst of a fixed initial mass.  The actual mass {\it still\/} in stars declines as matter is returned to the ISM; with the specific IMF and metallicity we have chosen, a stellar population loses $\sim12$, $\sim28$, $\sim 39$, and $\sim 49$ percent of its mass after 10~Myr, 100~Myr, 1~Gyr, and 10~Gyr, respectively.  However, this process has not been included in the N-body simulations\footnote{Preliminary experiments by JEB suggest that a Monte-Carlo scheme for returning stellar particles to the gas phase is workable, but further development of this scheme is beyond the scope of this paper.}; particles representing stellar components have fixed masses, and the total number of such particles is either fixed (without star formation) or monotonically increasing (with star formation).  The question is how to reconcile population synthesis, which includes mass loss, with N-body simulations, which do not.  We chose to take the masses of our stellar particles ``at face value'' when assigning luminosities; we therefore correct BC03's tabulated luminosities upward so that they represent a population with a {\it constant\/} stellar mass.  As indicated by the mass-loss figures given above, this correction is small for relatively young stellar particles but becomes significant for older ones.  For example, when evaluating the luminosity of a bulge particle with an age of 10~Gyr, we multiply the tabulated luminosity by a factor of $(1 - 0.49)^{-1}\simeq1.96$.  Since each stellar particle, including those formed during the simulation, has a definite birth-date, it is straightforward to apply this correction on a particle-by-particle basis when assigning luminosities.

The surface brightness along each line of sight, $I_{X}$, is then given by the standard equation of radiative transfer with no scattering:
\begin{equation}
\frac{dI_{X}}{dz}=j_{X}-\rho_{g}\kappa_{X}I_{X},
\end{equation}
\noindent where $z$ is the distance traveled in the direction of the ray to the observer, $\rho_{g}$ is the density of the gas and $\kappa_{X}$ is the opacity.  We define the opacity according to the values of Milky Way so that a gas surface density of $2.0\times10^{7}$~M$_{\sun}$~kpc$^{-2}$ yields an extinction of $A_{V}=1.00$, $A_{U}=1.53$, and $A_{B}=1.32$.  Total magnitudes and colors are obtained by summing the surface brightness of the entire galaxy at each waveband.

\section{Morphology and Kinematics\label{morkin}}
The orbital geometry of our simulations is largely based on HM95.  The initial orbits were parabolic, with an pericentric separation of $2.4\,R_{disk}$, or $r_{p}=$ 4.4~kpc.  We found that the angles specifying the orientation of the progenitor disks in HM95 were slightly ambiguous; HM95's definition of the argument of pericenter, $\omega$, differs from the definition given by \citet{tt}.  Fortunately we were able to match the geometry of the initial conditions based on the projections on the $x$-$y$ and $y$-$z$ planes \citep[$cf.$~Figure~AVIII-1 in][]{hb95}, obtaining inclination, $i$, and arguments, $\omega$, of $(i_{NW},\omega_{NW})=(-40,90)$ and $(i_{E},\omega_{E})=(70, 130)$, for the disks giving rise to the Northwest tail and the Eastern tail respectively.  In the rest of this paper, this combination of pericentric separation and orbital angles is referred to as the {\em HM95 setup} for convenience.  We also experimented with several combinations of different angles and pericentric separations close to this setup.  

\begin{figure}
  \centering
   \includegraphics[width=3.2in,angle=-90]{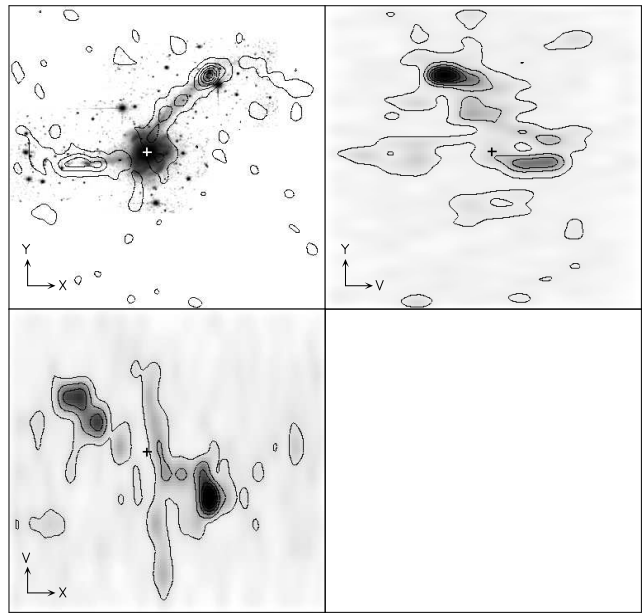}
   \caption{A snapshot image of \idkit$1.0$ \citep{bh}.  Top Left: Optical image of NGC~7252 overlaid with H\,{\small I} contours (HGvGS94).  Top Right: $Y-V_{z}$ position-velocity diagram.  Bottom Left: $V_{z}-X$ position-velocity diagram (HM95).}
   \label{idkt1}
\end{figure}

\citet{bh} have developed a software package, \idkit$1.0$, which allows rapid exploration of the encounter parameter space to reproduce the observed morphology and kinematics of interacting disk galaxies.  In the following we describe the steps of our matching process, which utilize this interactive software extensively.  Figure~\ref{idkt1} shows an image from the software, with the observed image and H\,{\small I} velocity data of NGC~7252 projected on each plane.  The software color codes the particles according to their galaxy of origin, and displays them according to their projected velocities and positions.  While the optical image and gas distribution show the observed morphology to be matched, the aligned panels of gas velocity provide strong constraints on the dynamics of the galaxy.  

First, an initial merger model was constructed with the HM95 setup, and self-consistent simulations incorporating both star formation rules were run based on this setup.  For this simulation, we used the interactive display of \idkit  to find a best match to the observed morphology and kinematics.  We then calculated the photometric properties of this model and compared them with the observed values, however we found that this model cannot explain the ages of young globular clusters observed by SS98, and neither the predicted star formation histories nor the photometric properties from this setup appeared consistent with the observations.

Thus the next step was to try different angles around the HM95 setup as well as different pericentric separations.  As described in \citet[][$cf.$~their Figure~3]{bh}, in addition to allowing the exploration of the parameter subspace with a fixed set of initial conditions, \idkit also provides instantaneous access to the outcome of different encounter geometries.  We used this capability by changing the pericentric separation, $r_{p}$, examined \idkit simulations with different choices of ($i$, $\omega$) around the HM95 setup, and found possible combinations of ($i$, $\omega$) and viewing directions which matched the morphology and kinematics of NGC~7252.  We then ran self-consistent simulations incorporating both star formation rules based on this setup, and compared the star formation histories as well as the photometric properties to the observed values.  If the results were inconsistent with the observations, this process was repeated with another choice of $r_{p}$ and the corresponding combination of ($i$, $\omega$), until a plausible match to the observational properties was found.  

The range of $r_{p}$ that we explored was 1.4 to 3\,$R_{disk}$, or $r_{p}=$ 2.6~kpc to 5.5~kpc.  The inclination angle $i_{NW}$ was fixed as $i_{NW}=-40$ in all the experiments, while the other angles were varied in the range of $\omega_{NW}=66$ to $90$, $i_{E}=60$ to $70$, and $\omega_{E}=130$ to $137$.  We emphasize here that, within the restrictions of time and machine power as well as the limitations of our numerical technique, an exact match to NGC~7252 is not possible; our goal is to obtain a star formation model that approximates the observational properties and provides reasonable predictions of NGC~7252's star formation history.  In the following we present our final best-fit model and discuss some of the discrepancies between our model and the observations.

\begin{figure}
  \centering
  \includegraphics[width=3.3in]{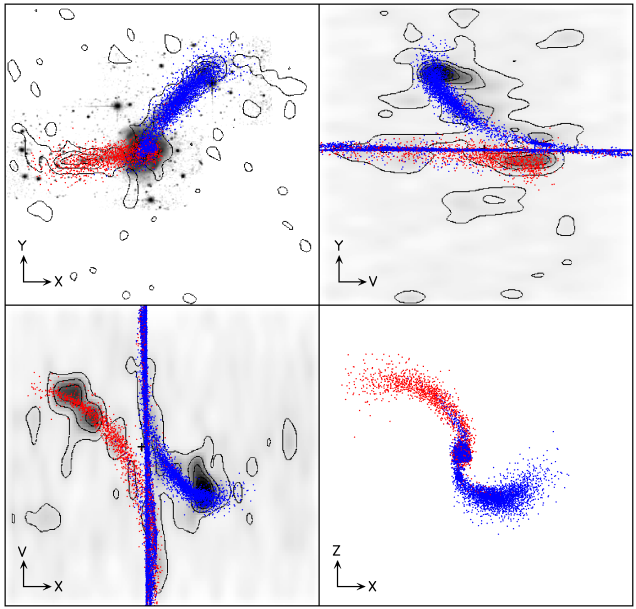}
  \caption{The final best-fit model of NGC~7252, displayed as color-coded particles on top of the observed data, according to their galaxy of origin.  Red particles relate to the progenitor galaxy that formed the E tail, while blue particles relate to the galaxy that formed the NW tail.  Bottom Right: Modeled NGC~7252 viewed from the projected $X-Z$ plane, where $Z$ is the direction along the line of sight.}
  \label{r159-idkt}
\end{figure}

The best-fit model assigns the two progenitor galaxies orbital angles of $(i_{NW},\omega_{NW})=(-40,66)$ and $(i_{E},\omega_{E})=(60,137)$ and adopts a pericentric separation of $r_{p}=1.92\,R_{disk}=$ 3.5~kpc.  The best match to the present occurs at $\simeq620$~Myr after periapse and is shown in Figure~\ref{r159-idkt}.  The overall morphology and kinematics of our simulation agree very well with the optical image and H\,{\small I} data.  In our best fit, the system is observed $\sim60\degr$ above the orbital plane and $\sim70\degr$ behind the line connecting the galaxies at orbital periapse.

As in other simulations of gas-rich merger remnants \citep[e.g.][]{b02}, our model contains a rotating gas disk superficially similar to the one observed in NGC~7252 \citep{s82}.  However the disk in our model appears nearly edge-on at the time of best match, of which observations indicate an inclination of $\sim40\degr$.  The simulated disk seems to be precessing in the gravitational potential of the remnant, with a rate which presumably depends on the details of the model, including the initial mass profiles of the progenitors and the smoothing of the force calculations \citep{b02}.  The orientation of the disk at the instant of best match might be improved by adjusting these details while preserving the overall choice of initial orbit, disk orientations and viewing angles; however this is beyond our resources.


\section{Star Formation History of NGC~7252}
\subsection{Interaction-Induced Star Formation}
\begin{figure}
  \centering
  \includegraphics[height=3.3in,angle=-90]{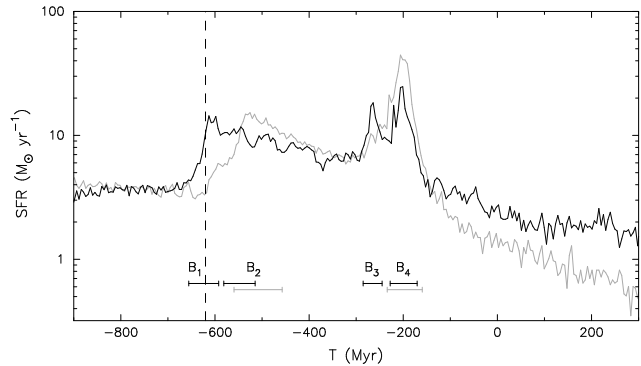}
  \caption{Global star formation history (in M$_{\sun}$~yr$^{-1}$) of NGC~7252 using two star formation rules.  Periapse occurred at $T\simeq-620$~Myr, indicated by the dashed line, and the best-fit time is at $T=0$.  Black and grey lines represent the SFR using the \si and the \dd star formation rules individually.  Horizontal lines at the bottom indicate the duration of each burst described in the text.}
  \label{sfhl-label}
\end{figure}

\begin{figure*}
  \centering
  \includegraphics[width=3.5in]{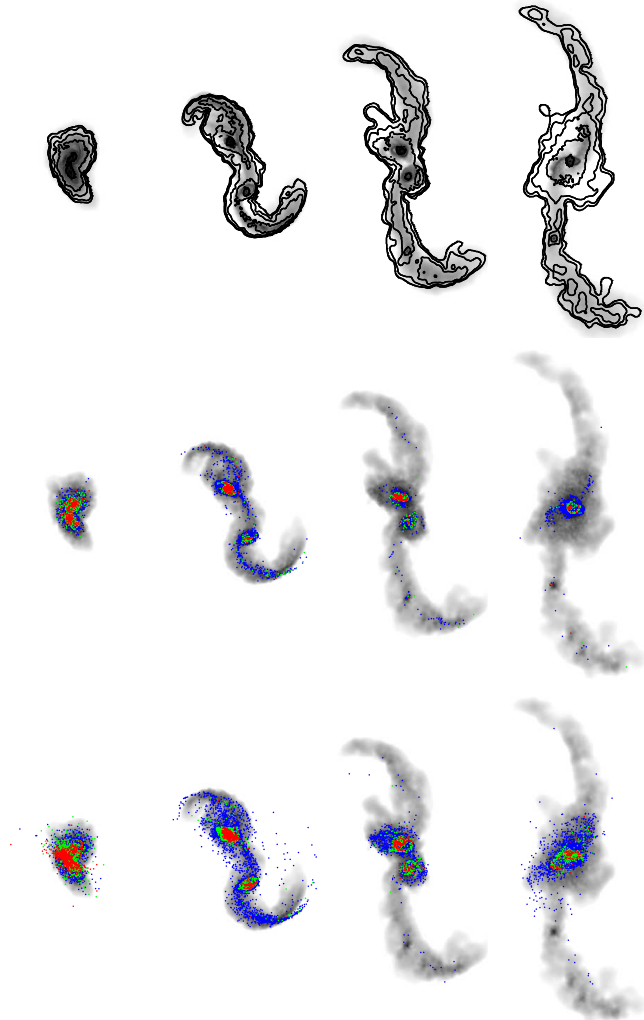}
  \caption{Dynamical evolution and star formation in simulations of NGC~7252 in a progression of time steps equal to 155~Myr at $T$~$\simeq-620$, $-465$, $-310$ and $-155$~Myr (from left to right).  Top Row: Gray scales show the gas distribution overlaid with contours as the distribution of old disk stellar populations.  No star formation is included.  Middle and Bottom Rows: In each image old stellar distribution is shown in halftone while star formation is shown as points in the \dd (middle) and the \si simulations (bottom).  Red points are stellar particles that have ages $\tau \la10$~Myr, green have $\tau \la40$~Myr, and blue have $\tau <155$~Myr.}
  \label{seq}
\end{figure*}

Figure~\ref{sfhl-label} shows the overall star formation histories predicted using the two different star formation models.  In this section we denote $T = 0$ as the best-fit time, i.e.~the present; thus the periapse occurred at $T\simeq-620$~Myr.  In the early stage of interaction, the two galaxies fall towards each other and interpenetrate at periapse.  This produces a prompt starburst in the \si simulation, where star formation quickens slightly before periapse, peaking at the time of closest approach.  In the \dd simulation, the star formation rate slowly rises after periapse as gas interacts within the perturbed disks and is driven into central regions.  The increasing central gas density eventually yields a broad peak in star formation rate at $T\simeq-540$~Myr.  Due to orbital decay, the two galaxies return for a second passage and collide again at $T\simeq-260$~Myr.  This results in another sharp peak in the \si simulation, where fresh gas constantly returns to galaxies from the tails and settles in the disk plane.  In the \dd simulation, following the pattern of the first passage, the star formation rate does not increase significantly until the gas is driven in towards the center.

At $T\simeq-215$~Myr the two galaxies merge.  In the \dd simulation a sudden increase in gas density in the nuclear regions results a strong and somewhat prolonged burst of star formation rate, which lasts to $T\simeq-155$~Myr.  In contrast, the \si simulation shows a much more prompt but brief starburst.

Figure~\ref{seq} shows the dynamical evolution of gas, old disk stars and interaction-induced star formation, in a progression of time steps equal to 155~Myr.  The top row shows the distribution of gas and the old disk stellar populations without any star formation, while the middle and bottom rows contrast star formation in the \dd and \si simulations.  At each time in the \dd simulation, most ongoing star formation, indicated by the red points, occurs at the central regions of the galaxies.  On the other hand, at the first passage in the \si simulation, the starburst happens along the interface where the two disks collide, and the star formation extends to the outskirts of the gas disks.  At $T\simeq-465$~Myr some star formation products (blue and green points) populate the gas poor regions beyond the tip of the lower tail.  These populations formed from the impinging gas, which is converted into stars so promptly that it still retains much of its momentum and so populates the gas poor regions.  Overall, the star formation has a wider spatial distribution in the \si than in the \dd simulation.  This larger spatial extent of star formation in \si model is a corollary of the earlier onset of activity in such models, since the gas is more widely distributed at earlier times \citep{b04}.

\begin{figure*}
  \centering
  \includegraphics[width=6.5in]{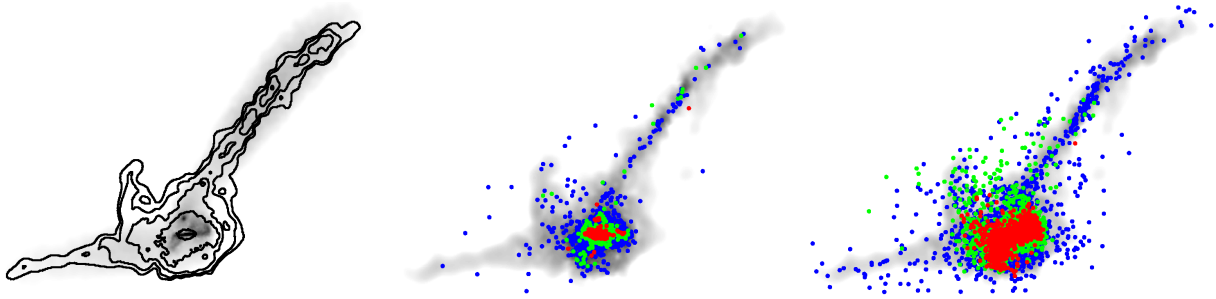}
  \caption{Best match of the simulation to the present configuration of NGC~7252.  Left: Gray scales show the gas distribution overlaid with contours as the distribution of old disk stellar populations.  Middle and Right: In each image old stellar distribution is shown in halftone and star formation is shown as points in the \dd (middle) and the \si model (right).  Red points are stellar particles that have ages $\tau <155$~Myr, green have 155--310~Myr and blue have 465--620~Myr.  
Same number of points are plotted in both simulations.}
  \label{sky}
\end{figure*}

Figure~\ref{sky} shows matches to the present configuration of NGC~7252 for the two star formation prescriptions.  The spatial extent of past star formation in the simulations is reflected in the distribution of star formation products at the present time.  Here the stellar particles that formed during the merger are color coded according to their ages.  The blue particles have ages of 465--620~Myr, corresponding to the products of the starburst occurred at the first passage ($T\simeq-620$~Myr).  The green particles have ages of 155--310~Myr, and the red particles have ages less than 155~Myr, reflecting regions of more recent star formation.  Note that in both simulations the same total number of particles are plotted.  In both cases, most of the on-going star formation is now confined to the central regions of the remnant body.  However products of star formation are found throughout the tails and the remnant, especially in the \si simulation.  The prompt starburst occurred at the first passage in the \si simulation produces a wide distribution of star formation throughout the whole galaxy, while in the \dd simulation the first passage did not induce a prompt burst of extended star formation, and the past star formation products are far less distributed outside the central body.

\begin{table}
  \centering
  \caption{Median radii (in kpc) for the burst populations.}
  \label{medradius}
  \begin{tabular}{@{}lccccc}
  \hline
  &B$_{1}$&B$_{2}$&B$_{3}$&B$_{4}$&all\\
  \hline
  {\it Density-dependent}&...&0.13&...&0.07&0.64\\
  {\it Shock-induced}&4.32&1.32&1.29&0.17&2.11\\
  \hline
  \end{tabular}
\end{table}

To quantify the spatial distribution of new stellar particles, we define four distinct starbursts in the \si simulation, as shown in Figure~\ref{sfhl-label}: `B$_{1}$' (ages of 590--655~Myr), `B$_{2}$' (515--580~Myr), `B$_{3}$' (245--285~Myr), and `B$_{4}$' (170--225~Myr).  These bursts are associated with the maxima in star formation rate.  In the \dd simulation, `B$_{2}$' (455--560~Myr) is defined as the broad star formation feature occurred after the first passage, and `B$_{4}$' (160--230~Myr) is related to the star formation peak at merging.  Table~\ref{medradius} lists the median 
radii in kpc for the burst populations, measured at the present time.  The last column ``all'' in the tables refers to the stars formed throughout the simulations from $-930$~Myr to present.  The center of the galaxy is computed by locating the the minimum of the potential well of the old bulge particles.  In the \dd simulation, most burst populations, and indeed most stars formed throughout the encounter are concentrated within the central kpc.  On the other hand, the burst population formed at the first passage in the \si simulation extends more than 4~kpc from the center.  Overall, the radial extent of stars formed in the \si simulation is nearly $4$ times larger than the extent of those formed in the \dd case.  The total initial gas mass of each simulation is $9.4\times10^{9}$~M$_{\sun}$.  At present, the \dd simulation produces, in total, $9.1\times10^{9}$~M$_{\sun}$ of stellar particles and \si simulation produces a comparable amount of $8.7\times10^{9}$~M$_{\sun}$.

\subsection{Distribution of Star Clusters}
\begin{figure*}
  \centering
  \includegraphics[width=3.5in,angle=-90]{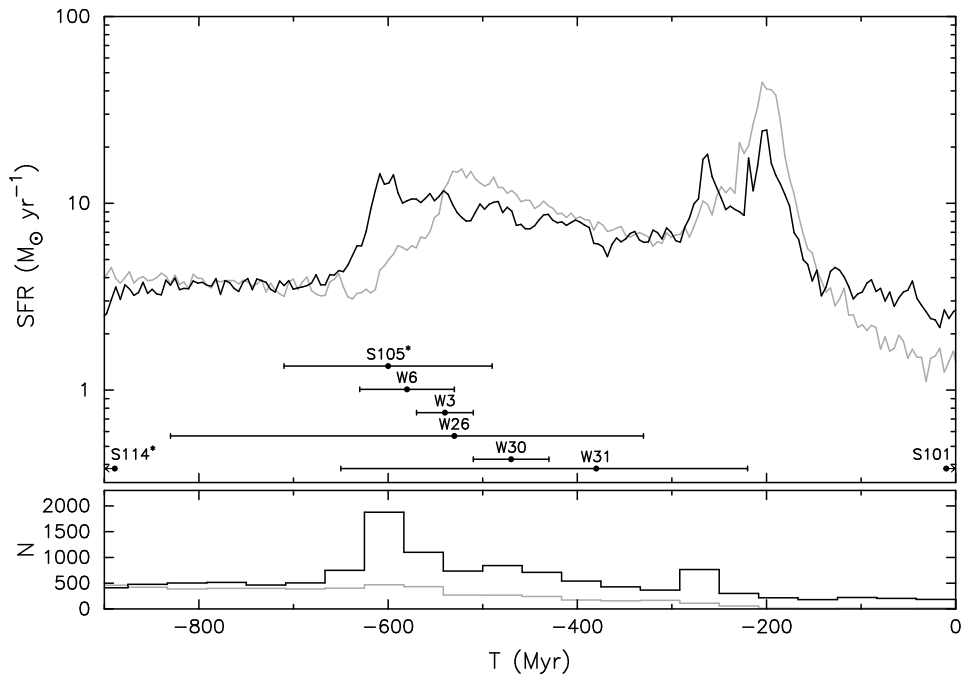}
  \caption{Comparison of age distribution of stellar populations.  Top Panel: Global star formation history (in M$_{\sun}$~yr$^{-1}$) of NGC~7252 shown from $-900$~Myr to present ($T=0$).  Color indications are the same as Figure~\ref{sfhl-label}.  Cluster ages from SS98 are plotted as dots with their uncertainties.  Cluster S101 has an age upper limit of $10$~Myr and cluster S114 has an age of $\sim1$~Gyr.  Note that cluster S105 and S114 each has another possible age of $\sim200$ and $\sim40$~Myr.  Bottom Panel:  Histograms of number of stellar particles formed in the simulation, located within 3--11~kpc from the center, measured at present time.}
  \label{sfhlt15x_cahist2}
\end{figure*}

SS98 identified and measured ages of 7 proto-globular clusters and 1 young stellar association in the remnant.  Their IDs and ages are compared with the simulated star formation histories of NGC~7252 in Figure~\ref{sfhlt15x_cahist2}.  Among the globular clusters, six of them lie between projected distances of 3--11~kpc from the center of NGC~7252 and have ages of $\sim400$--$600$~Myr, indicating that they formed early in the recent merger.  Some of the cluster ages have a wide uncertainty; both of our simulations successfully reproduce the range of cluster ages, although clusters W6 and S105, with ages of $580$ and $600$~Myr, are more consistent with the prompt starburst at first passage in the \si simulation.  

The spatial distribution of the observed clusters, however, strongly discriminates between our models.  To compare with the observed locations of these clusters (3--11~kpc from the center), the bottom panel of Figure~\ref{sfhlt15x_cahist2} shows the age distribution of stellar particles located within this annulus in our simulations.  The \dd simulation, indicated as the grey line, produces a gradually declining distribution of ages, with almost all interaction-induced star formation within the very central regions.  In contrast, the \si simulation produces many more stellar particles in this annulus and a sharp peak around the first passage $\simeq-620$~Myr, showing that star formation occurs in more dispersed regions away from the center and that a high portion of stellar particles within this annulus formed during the starburst in the first passage.  This result appears consistent with the distribution of ages observed in SS98, suggesting that shocks can be an important trigger of formation of these clusters.

\begin{figure}
  \centering
  \includegraphics[width=3.3in]{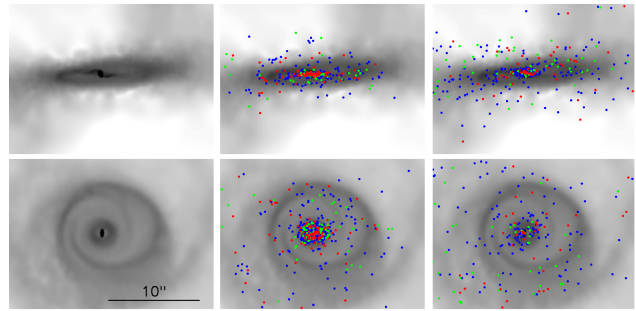}
  \caption{View of the central disk of our best-fit model from the sky (Top) and the orbital plane (Bottom).  Left: Gray scales show the gas distribution.  Middle and Right: In each image the gas disk is shown in halftone and star formation is shown as points in the \dd (middle) and the \si model (right).  Red points are stellar particles that have ages $\leq 10$~Myr, green have $10 < \tau < 20$~Myr and blue have $20 < \tau < 50$~Myr.  Same stellar particles are shown in both views for each model, and the same total number of points are plotted in both simulations.}
  \label{sky-cdisk}
\end{figure}

\citet{m97} found that the youngest stellar populations in NGC~7252, with ages of $\sim10$~Myr, lie within 2.0~kpc in the central disk.  In our simulations, $93\%$ of the stellar particles with ages $\leq10$~Myr in \dd and $84\%$ in the \si models are formed within 2.0~kpc from the center.  A further inspection of the central region is shown in Figure~\ref{sky-cdisk}.   The top row shows the view of the central disk of our best-fit model from the sky and the bottom shows the view from the orbital plane.  Despite the inconsistent orientation of our disk mentioned in Section 3, the size is consistent with the observed scale of about 2.5~kpc in radius \citep{s82}.  The red points in Figure~\ref{sky-cdisk} represent the stellar particles that have ages $\leq10$~Myr; in both \dd and \si models most of these young stellar particles are formed within the disk.

\section{Photometric Properties of NGC~7252}
\begin{figure*}
  \centering
  \includegraphics[height=5.5in,angle=-90]{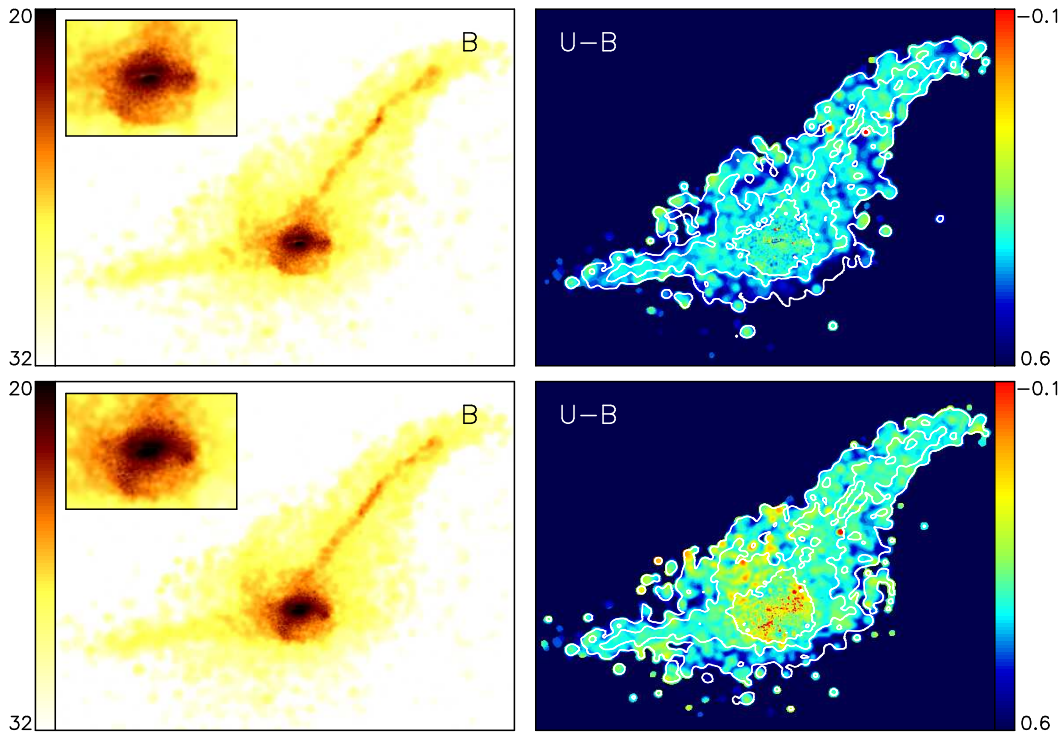}
  \caption{Comparison of simulated $B$ and $U-B$ images of NGC~7252 using \dd (Top) and \si (Bottom) star formation rule.  In the $B$ images, grayscale represents surface brightness ranging from $\mu_{B}=$ 20 (dark brown) to 32 (white)~mag~arcsec$^{-2}$.  In the $U-B$ images grayscale ranges from $U-B=-0.1$ (red) to 0.6 (blue).  Contours represent $\mu_{B}$ with levels of $26.0$, $28.0$ and $30.0$~mag~arcsec$^{-2}$.}
  \label{colorimage}
\end{figure*}

Figure~\ref{colorimage} shows the simulated $B$ and $U-B$ images of \dd and \si models which best fit the present configuration.  The $B$ band images of both models show a dust lane across the central regions.  There are ripples and loops around the remnant in both models, and star forming regions in the tails are seen as brighter clumps in the images.  Since $U-B$ traces young and massive stars, the $U-B$ images indicate where the most violent star formation is.  

Table~\ref{photcolors} lists the observed and modeled $M_{B}$ and colors of NGC~7252; magnitudes and colors computed without extinction are included in the parentheses.  The colors from the \si model seem a closer match to the observed values.  Integrated colors of the \dd model appear to be redder than those of the \si model.  This can be considered a direct result of the star formation prescription since star formation tends to concentrate in regions of high gas density in the \dd model, and thus gives a higher extinction and much redder colors than the \si model.  For example, in \dd model extinction causes $M_{B}$ to be $1.51$~mag fainter and the colors to be $\ga0.2$~mag redder.  However, the total $M_{B}$ of both models are still fainter than the observed values.  

Despite the good match to the morphology of our best-fit model, the tails in our model seem to be too faint compared to the observations.  Table~\ref{tailprop} lists $B$ magnitudes and $B-R$ colors for the tails from the observations and from our best-fit model.  HGvGS94 mentioned that the tails are $0.2$--$0.5$~mag bluer in $B-R$ than the outer regions of the main body, however they did not give specific values for the tails.  Averaging the $B-R$ colors of the outer regions of the main body (region A through C and H through M in HGvGS94) gives a value of $1.3$, so the average $B-R$ color of the tails is about $0.8$--$1.1$.  The values from our models are obtained by measuring our simulated images within the same areas of the tails from HGvGS94.  In both models, we find that the tails are fainter by about $2$~mag in $M_{B}$ than the observed value and have redder $B-R$ colors.  

Many factors could contribute to this result and it is beyond our scope to investigate each in detail.  Nevertheless it seems possible that the main reason is a lack of star formation in the tails, as reflected in the redder colors of our simulations.  The lack of star formation could be due to insufficient material in the tails, or to an incomplete description of star formation in the tails.  A simple check of the gas in the tails helps to distinguish between these possibilities.

\begin{table*}
 \centering
  \caption{Photometric properties (in mag) of NGC~7252.}
  \label{photcolors}
  \begin{tabular}{@{}lllll}
  \hline
  &$M_{B}$&$U-B$&$B-V$&$V-R$\\
  \hline
  Observed & -21.22 & 0.17 & 0.66 & 0.74\\
  {\it Density-dependent} & -20.08 (-21.59) & 0.24 (0.02) & 0.70 (0.41) & 0.55 (0.39)\\
  {\it Shock-induced} & -20.68 (-21.54) & 0.16 (-0.03) & 0.60 (0.42) & 0.50 (0.39) \\
  \hline
  \end{tabular}
\end{table*}

In our best-fit simulations, there is about $4\times10^{8}$~M$_{\sun}$~of gas in the NW tail and about $1.3\times10^{8}$~M$_{\sun}$~in the E tail (Table~\ref{tailprop}); in round numbers, our modeled tails have $\sim20\%$ of the gas observed in the real tails.  This deficiency very likely occurs because our gas particles were initially distributed like disk particles.  If we were to use a more realistic gas distribution for our progenitor galaxies, i.e.~typical late-type gas disks with radii extending to about twice those of the stellar disks, a much larger gas mass in the tails could be achieved.

However, even if the gas content of our tails could be increased by a factor of $\sim5$ to match the observations of the NW tail, the star formation associated with this gas content is still probably insufficient.  In the NW tail, about $9\times10^{4}$ and $2\times10^{5}$~M$_{\sun}$ of stars were born in the past $100$~Myr in the \dd and the \si model respectively.  Thus we estimate the present star formation rate as $0.002$~M$_{\sun}$~yr$^{-1}$ in the NW tail, assuming a constant star formation rate for the past $100$~Myr.  According to equation (\ref{sfp}), if we scale the gas content in the modeled tails to the observed amount of gas, the maximum star formation rate would become about $0.01$--$0.02$~M$_{\sun}$~yr$^{-1}$.  This increase of star formation rate would bring the total $B$-band luminosity in the NW tail to about $0.5$--$1\times10^{9}$~L$_{\sun}$ for each model, assuming a Salpeter initial mass function and solar metallicity \citep{l99}.  The resulting $M_{B}$ would become $-16.4$ and $-17.3$~mag for the \dd and \si model respectively, which are still fainter than the observed magnitude of $-18.1$.  

The above calculations show that even if we match the observed amount of gas in both tails, our models still would not reproduce the observed brightness of NGC~7252's tails.  In fact, the tails in our models contain enough gas to account for the observed tail luminosities, if we consider a constant star formation rate in the tails since the first passage.  For example, the NW tail has sufficient gas to sustain a star formation rate of $\sim0.72$~M$_{\sun}$~yr$^{-1}$ for the past $\sim620$~Myr.  The stars formed would produce a $B$-band luminosity of $\sim6\times10^{9}$~L$_{\sun}$, increasing the magnitude of the NW tail to $M_{B}\simeq-19$~mag, or about one magnitude brighter than observed.  

These results strongly indicate that our tails appear fainter and redder than the observation due to an incomplete description of star formation.  Clearly, the star formation rules could be improved.  In low density environment such as tails, \dd rules may underestimate the star formation rate based on its dependence of gas density.  On the other hand, \si rule might not be useful since the tails are expanding kinematically.  One way to solve this problem is to modify the adopted star formation rules.  For example, another mode of star formation could be added to the present form, in which the star formation rate is proportional to the gas density to the order of unity, so that $\dot{\rho}_{\ast} \propto C_{\ast}~\rho_{g}^{n}~\dot{u}^{m} + C_{\circ}~\rho_{g}$.  This additional term would then represent a finite chance \citep[$=C_{\circ}~\Delta$$t$, where $\Delta$$t$ is each simulation time-step; see][]{b04} for gas to form stars regardless of the circumstances in the simulations.

\begin{table*}
 \centering
  \caption{Properties of the tails.}
  \label{tailprop}
  \begin{tabular}{@{}lccccccc}
  \hline
  &&NW tail&&&&E tail&\\
  \cline{2-4}
  \cline{6-8}
  &$M_{B}$&$B-R$&H {\small I} Mass&&$M_{B}$&$B-R$&H {\small I} Mass\\
  &(mag)&(mag)&($10^{8}$~M$_{\sun}$)&&(mag)&(mag)&($10^{8}$~M$_{\sun}$)\\
  \hline
  Observed$^{a}$ & -18.09 & $0.8-1.1$ & 20.8$\pm$0.5 && -16.84 & $0.8-1.1$ & 11.0$\pm$0.4\\
  {\it Density-dependent} & -15.96 & 1.32 & 4.5 && -14.60 & 1.37 & 1.4\\
  {\it Shock-induced} & -16.12 & 1.25 & 4.2 && -14.87 & 1.27 & 1.3\\
  \hline
  \end{tabular}\\
  \medskip
  $^{a}$Values adopted or calculated from \citet{hb94}.
\end{table*}

\section{Summary}
We describe new simulations of the galactic merger NGC~7252 with gas dynamics and star formation included.  In our models NGC~7252 is formed by the merger of two similar gas-rich disk galaxies which fell together with an initial pericentric separation of $\sim1.9$~disk scale lengths about 620~Myr ago.  Starting with plausible models for the pre-encounter disks, our simulation produce fairly well matches with the observed morphology and kinematics.  We emphasize that rather than a perfect reproduction, our goal is to present a plausible dynamical model of NGC~7252, which provides insight into the star formation history and matches the photometric properties; this nonetheless represents a new and significant step beyond pure dynamical modeling.  

Following the study of NGC~4676 by \citet{b04} and \citet{c07}, we consider two star formation mechanism in our simulation and use the ages of young globular clusters from \citet{ss} to constrain the trigger of star formation.  \citet{b04} suggested that \dd star formation rule gives an incomplete description of large scale star formation in interacting galaxies.  In our simulations of NGC~7252, \dd star formation happens mostly at gas rich regions concentrated in galactic centers throughout the simulation, whereas more products of past star formation in \si model are distributed at wider projected distances and in tails.  The total mass born in bursts are on the same order in each model, yet the median projected radius of the burst populations is 0.6~kpc in \dd model and 2.1~kpc in \si model.  This supports the suggestion in \citet{b04} that \si star formation may provide a better match to the extended distributions of young clusters in NGC~7252 and other merger remnants. 

When compared with observations, the star formation histories of both models successfully reproduce the range of ages spanned by the observed clusters.  Those that have ages of $\sim570$--$600$~Myr, in particular, are likely to form during the starburst at first passage, and this seems consistent with the prediction of the \si simulation.  Within the projected annulus where the observed clusters lie, the age distribution of stellar particles formed in the \si simulation shows a peak around such age, while in the \dd simulation the age distribution is flat and declining.  Again, almost all interaction-induced star formation occurred in the very central regions for the \dd model, and the \si model is more likely to explain the concentration of ages of clusters observed in SS98.

We also describe a method to incorporate simple stellar population synthesis models into our dynamical simulations in order to compare the predicted photometric properties with the observed values.  Beyond the attempt of matching the morphology as in many previous simulations of interacting galaxies, we also try to match the photometric properties.  While the overall magnitudes and colors are consistent with the observed values, the tails seem to be fainter and redder than observed.  We argue that this lack of star formation, reflected by the redder colors, is due to an incomplete description of star formation, rather than insufficient material in the tails.  An additional mode of star formation which is independent of large-scale conditions could help to solve this problem.

Finally we discuss some future directions for making our star-forming models more realistic, which could help improve the modeled photometric properties.  First, as discussed in Section \ref{mphot}, our models do not incorporate the gradual mass return from evolving stellar populations.  In order to incorporate population synthesis models to produce luminosities, we have to make adjustments which are not entirely straightforward.  Moreover, recycled gas might help boost the star formation rate throughout the last stages of the merger, improving the luminosities and colors of our models of NGC~7252.  

Another concern regarding the colors of the merger remnant tails comes from the colors of the progenitor disks.  Disk galaxies generally become moderately but systematically bluer outward from the bulge, and this effect is caused by the age and metallicity gradients in the stellar populations as well as internal dust extinction \citep[e.g.][]{wk87, BdJ00}.  In order to improve our modeling of the progenitor galaxies, we could include radial gradients of age and metallicity in the initial stellar disks.  This could influence the photometric properties of the remnant tails, since they mostly come from the outer parts of the progenitor disks.  With regard to our models of NGC~7252, it might also help to improve the luminosities and colors of the tails.  However, unless star formation in the outer regions and tails can be sustained throughout the merger process, the outer populations will fade fast after the encounter, reducing the effect of age gradients on tail properties.

\section*{Acknowledgments}
We thank our referee Chris Mihos for many useful suggestions on the manuscript and Francois Schweizer for helpful comments.  We also thank the Kyoto University Astronomy Department and the chairman Tetsuya Nagata for support and hospitality.  L.-H. C. gratefully acknowledges the support from the JPL Research Support Contracts RSA 1282612 and 1298213 (PI: D. B. Sanders), NSF AST-0707911 (PI: F. Bresolin), and the IfA Research Support Funds.


\clearpage

\clearpage

\end{document}